\documentclass[conference]{IEEEtran}
\IEEEoverridecommandlockouts
\usepackage{xspace,amsfonts,amsmath,amssymb,epsfig,syntonly}
\usepackage{cite}

\newtheorem{proposition}{\hspace{-11pt}\bf Proposition}

\newtheorem{remark}{\hspace{-11pt}\bf Remark}

\begin{document}
\title{ Spectral Efficiency of the Cellular Two-Way  Relaying with Large Antenna Arrays}

\author{
\IEEEauthorblockN{Zhaoxi Fang}
\IEEEauthorblockA{Dept. of Commun.   Eng.\\
Zhejiang Wanli University\\
Ningbo, China}
\and
\IEEEauthorblockN{Xin Wang}
\IEEEauthorblockA{Key Laboratory of EMW Information (MoE)\\
Dept. of Commun. Sci. \& Eng., Fudan University\\
Shanghai, China}
\and
\IEEEauthorblockN{Xiaojun Yuan}
\IEEEauthorblockA{School of Info. Sci. \& Tech.\\
ShanghaiTech University \\
Shanghai,  China}

\thanks{Work in this paper was supported by   the Natural Science Foundation of China under Grant No. 61401400, the National Key Technology R\&D Program of China under Grant No. 2013BAK09B03, the Innovation Team of Ningbo Science and Technology Bureau under Grant No. 2013B82009, and the Natural Science Foundation of Ningbo under Grant No. 2013A610121.}
}

\markboth{Draft \today}%
{Zhaoxi Fang  }

\maketitle 

\begin{abstract}
This paper  considers a  multiuser cellular two-way  relay  network (cTWRN) where multiple users exchange information with a base station (BS) via a relay station (RS). Each user is equipped with a single antenna, while both the BS and the RS are equipped with a very large antenna array. We investigate  the performance of the cTWRN with amplify-and-forward (AF) based physical-layer network coding,  and derive closed-form expression  for the asymptotic spectral efficiency when both the number of antennas at the BS and the RS grow large.  It is shown that the noise propagation of the non-regenerative relaying protocol  can be greatly suppressed, and the AF relaying scheme can approach the cut-set bound under certain conditions. We also investigate  the   performance of the AF relaying scheme under two power-scaling cases, and show that the transmit power of the BS and each user can be made inversely proportional to the number of relay antennas while maintaining a given quality-of-service. Numerical results are presented to verify the analytical results.

\end{abstract}

 \begin{keywords}
 Multiuser two-way relaying, large antenna array,  zero-forcing processing, spectral efficiency.
 \end{keywords}


\section{Introduction}
The   two-way relaying (TWR) technique has been generalized to   multiuser cellular networks recently \cite{Ding11,Sun12}, where multiple  users communicate with a base station (BS) via the help of a relay station (RS).  Several efficient amplify-and-forward (AF) and decoded-and-forward (DF) based  multiuser relaying protocols have been proposed for the cellular two-way relay network (cTWRN) \cite{Ding11,Chi12,Sun12,YangHJ12,Fang14}.  It was shown that multiuser two-way relaying has the potential to increase network throughput, extend the coverage, and reduce the power consumption of mobile terminals.

On a parallel avenue,  multiple-input multiple-output (MIMO) systems with  very large antenna  arrays (also known as ``massive MIMO'') have attracted a lot of research interest  recently \cite{Marzetta10,Ngo13,Rusek13}. It was shown in \cite{Ngo13} that an array of a very large number of antennas can reduce the effect of noise, average the small-scale fading, mitigate inter-user interference, and improve the spectral  efficiency with  simple linear signal processing techniques, such as matched-filtering (MF), or zero-forcing (ZF) precoding/detection. With these advantages, massive MIMO   has been considered as a   promising technique in future  cellular networks \cite{Rusek13}.

The spectral efficiency of the two-way relay network (TWRN)  can   be further improved with very large antenna arrays \cite{Suraweera13,Ngo14,Cui14}. In \cite{Suraweera13}, the authors investigated the power efficiency of a half-duplex multi-pair AF TWRN  with a large antenna array  at the relay.  It was shown that the transmit power of the sources can be greatly reduced while maintaining a given quality-of-service. For full-duplex multi-pair one-way relaying, the authors in \cite{Ngo14} showed  that  the loop interference can be significantly reduced  with massive receive antenna array  at the relay.

In this paper, we consider the cellular two-way relay network, where  multiple single-antenna users communicate  with a base station  via the help of a relay station.  Both the BS and the RS are equipped with very large antenna arrays.  We consider an amplify-and-forward physical-layer network coding (PLNC) scheme based on signal space alignment and zero-forcing processing, which will be  referred to  as SA-ZF in the following. Note that this scheme was originally proposed in \cite{Ding11} for a special case of cTWRN where   the  number of antennas at the BS and the RS are equal to the number of users, which has a relative low complexity as compared with DF-based relaying schemes \cite{YangHJ12,Fang14}. However, due to the noise propagation, there was at least 3 dB performance loss  as compared with the cut-set bound when both the BS and the RS are equipped with a finite number of antennas \cite{Ding11,Fang14}.

 In this paper, we analyze the spectral  efficiency of the   SA-ZF  relaying scheme  when both the number of   BS and   RS antennas grow large. It is  shown that the noise propagation effect   can be greatly suppressed, and the SA-ZF scheme with very large antenna arrays can approach the   cut-set bound under certain conditions. We also investigate   two power-scaling cases and show that the transmit power of  the BS and the users can be made inversely proportional to the number of relay antennas while maintaining a given quality-of-service. Numerical results are presented to verify the analysis. Both analytical and numerical results reveal that the cellular two-way relaying with large antenna arrays and simple linear processing   can mitigate the inter-stream interference (ISI), average the fast fading,   reduce the power consumption and improve the spectral efficiency of the cTWRN as in point-to-point massive MIMO systems.

The rest of the paper is organized as follows. Section II describes the cTWRN model and the AF-based SA-ZF scheme. Sections III   provides asymptotic performance analysis.  Numerical results are presented in Section IV, followed by the conclusions in Section V.

%
%

\section{System Model}
We consider a  multiuser cellular two-way relay network  as
shown in Fig. 1.   There is no direct link between the BS and the
users due to path-loss and shadowing, and the BS communicates
with $K$   users via a single relay station. The BS  is equipped with $N_B$ antennas, the RS   with  $N_R$ antennas, and each user  with single antenna.  The channels form the BS and the $k$-th user to the RS are denoted by $\boldsymbol{H}_{BR} \in {\cal C}^{N_R \times N_B}$ and $\boldsymbol{h}_{k,R} \in {\cal C}^{N_R \times1}$, respectively, while those from the RS to the BS and the $k$-user are denoted by  $\boldsymbol{H}_{RB} \in {\cal C}^{N_B \times N_R}$ and $\boldsymbol{h}_{R,k}^T \in {\cal C}^{1 \times N_R}$, respectively.  We assume all the nodes are half-duplex and each round of bidirectional data exchange consists of two phases.

In the first phase, the BS and all the users transmit to the RS simultaneously. Let  $\boldsymbol{s}_B \in {\cal C}^{K \times 1}$ denote the information symbol vector to be transmitted at the BS with $\mathbb{E}[\boldsymbol{s}_B \boldsymbol{s}_B^H] = \boldsymbol{I}_K$. The transmit signal of the BS is given by  
 \begin{equation}
 \boldsymbol{x}_B = \boldsymbol{F}_B \boldsymbol{s}_B,
  \end{equation}
where $\boldsymbol{F}_B \in {\cal C}^{N_B \times K} $ is the precoding matrix at the BS. The transmit signal of the $k$-th user is 
\begin{equation}
x_{U,k} = \sqrt{P_U} s_{U,k},
 \end{equation}
where $P_U$ is the transmit power and $s_{U,k}$ denotes the unit-power information symbol. In the first phase,  the received signal at the relay is given by \cite{Fang14}

\begin{equation}\label{eq.yR2}
\boldsymbol{y}_R  = \boldsymbol{H}_{BR} \boldsymbol{F}_B \boldsymbol{s}_B +  \sqrt{P_U} \boldsymbol{H}_{UR}  \boldsymbol{s}_{U} + \boldsymbol{z}_R,
\end{equation}
where $\boldsymbol{H}_{UR} = [\boldsymbol{h}_{1,R}, \ldots, \boldsymbol{h}_{K,R}]$,    $ \boldsymbol{s}_U = [s_{U,1}, \ldots, s_{U,K}]^T$, and $\boldsymbol{z}_R \thicksim {\cal CN}(0,\sigma^2 \boldsymbol{I}_{N_R})$ denotes the additive white Gaussian noise (AWGN) at the RS.  The channel matrix $\boldsymbol{H}_{BR}$ can be represented as $\boldsymbol{H}_{BR} = \sqrt{\ell_{B}} \tilde{\boldsymbol{H}}_{BR}$, where $\ell_{B}$ represents the path-loss from the BS to the relay, and $\tilde{\boldsymbol{H}}_{BR}$ represents the normalized small-scale fading coefficients matrix. Similarly, $\boldsymbol{H}_{UR} = \sqrt{\ell_{U}}  \tilde{\boldsymbol{H}}_{UR} $, where $\ell_{U}$ denotes the path-loss from the users to the relay, and $\tilde{\boldsymbol{H}}_{UR}$ denotes the small-scale fading coefficients matrix.

Based on the signal model in (\ref{eq.yR2}), linear precoding is employed at the BS, with the precoder given by
 \begin{equation}\label{eq.FB1}
\boldsymbol{F}_B = \alpha_B  \boldsymbol{H}_{BR}^{\dag} \boldsymbol{H}_{UR},
\end{equation}
where $\boldsymbol{H}_{BR}^{\dag} = \boldsymbol{H}_{BR}^H \left(\boldsymbol{H}_{BR} \boldsymbol{H}_{BR}^H \right)^{-1}$, and $\alpha_B $ is a constant  to meet the transmit power constraint at the BS: $   \mathbb{E} \left[ \text{tr} \left( \boldsymbol{x}_B \boldsymbol{x}_B^H\right) \right]  = P_B$.
With such precoding, the received signal at the RS is given by
 \begin{equation}\label{eq.yRSAZF}
 \boldsymbol{y}_R   =   \boldsymbol{H}_{UR}  \left( \alpha_B \boldsymbol{s}_B + \sqrt{P_U} \boldsymbol{s}_U \right)  + \boldsymbol{z}_R.
 \end{equation}

Upon receiving $\boldsymbol{y}_R$, the transmit signal at the RS  is regenerated as
\begin{equation}\label{eq.xR}
\boldsymbol{x}_R = \boldsymbol{W}_R \boldsymbol{y}_R,
\end{equation}
where $\boldsymbol{W}_R \in {\cal C}^{N_R \times N_R}$ is the linear processing matrix.

With an array of a very large number of antennas, the RS is able to null out the inter-stream inference   completely. The ZF receiving and   transmit precoding    matrix at the RS is given by
\begin{equation}\label{eq.WR1}
\boldsymbol{W}_R  = \alpha_R  \boldsymbol{H}_{RU}^{\dag} \boldsymbol{H}_{UR}^{\dag}.
\end{equation}
where $\alpha_R $ is chosen to meet the relay power constraint: $ \text{tr}\left( \mathbb{E} \left[\boldsymbol{x}_R \boldsymbol{x}_R^H  \right] \right)= P_R$, and $\boldsymbol{H}_{RU} = [\boldsymbol{h}_{R,1}, \ldots, \boldsymbol{h}_{R,K}]^T$.

\begin{figure}[t]
\centering
\includegraphics[width=3.5  in]{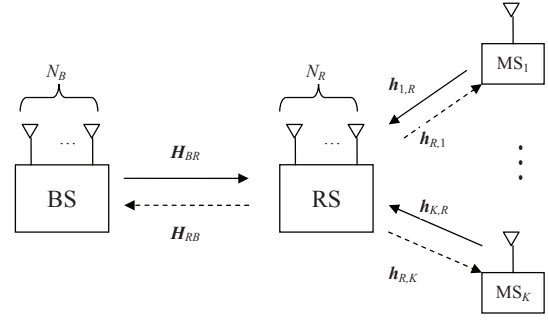}\\
 \caption{A cellular two-way relay  network with $K$ users.}
\end{figure}

In the second phase, the  relay broadcasts $\boldsymbol{x}_R$ to the BS and the users.
The received signal at the BS and the $k$-th user is respectively given by
\begin{equation}\label{eq.yB1}
\boldsymbol{y}_B = \boldsymbol{H}_{RB} \boldsymbol{x}_R + \boldsymbol{z}_B,
\end{equation}
and
\begin{equation}\label{eq.yMk1}
y_{U,k} = \boldsymbol{h}_{R,k}^T \boldsymbol{x}_R + z_{U,k},
\end{equation}
where   $\boldsymbol{z}_B \thicksim {\cal CN}(0,\sigma^2 \boldsymbol{I}_{N_B})$ and $z_{U,k} \thicksim {\cal CN}(0,\sigma^2)$ denote the AWGN at the BS and the $k$-th user, respectively.  Without loss of generality, we assume the channels are reciprocal that  $\boldsymbol{H}_{RB} = \boldsymbol{H}_{RB}^T$ and $\boldsymbol{h}_{R,k} = \boldsymbol{h}_{k,R}, \forall k$.

From (\ref{eq.yR2})-(\ref{eq.yB1}), the received signal at the BS can be expressed as
\begin{equation}\label{eq.yB2}
\begin{split}
\boldsymbol{y}_B &= \boldsymbol{H}_{RB} \boldsymbol{W}_{R} \boldsymbol{H}_{BR} \boldsymbol{x}_B + \boldsymbol{H}_{RB} \boldsymbol{W}_{R} \boldsymbol{H}_{UR} \boldsymbol{x}_U \\
& + \boldsymbol{H}_{RB} \boldsymbol{W}_{R}\boldsymbol{z}_R +  \boldsymbol{z}_B.
\end{split}
\end{equation}
Note that the first term on the right hand side of the above equation is the self-interference (SI) which is known by the BS and can be subtracted before signal detection. After removing the SI, a zero-forcing detection matrix $\boldsymbol{W}_B \in  {\cal C}^{N_B \times K}$ is applied to estimate the messages from the users as
\begin{equation}\label{eq.yB3}
\tilde{\boldsymbol{y}}_B  = \boldsymbol{W}_B^H (\boldsymbol{y}_B - \boldsymbol{H}_{RB} \boldsymbol{W}_{R} \boldsymbol{H}_{BR} \boldsymbol{x}_B),
\end{equation}
where the ZF receiving matrix $\boldsymbol{W}_B$ is given by
\begin{equation}\label{eq.WB}
\boldsymbol{W}_B =  \left(\boldsymbol{H}_{RU} \boldsymbol{H}_{RB} ^{\dag} \right)^H.
 \end{equation}
Substituting (\ref{eq.WB}) into    (\ref{eq.yB3}), the received $k$-th data stream at the BS is given by
 \begin{equation}\label{eq.yBk}
 \tilde{\boldsymbol{y}}_{B,k}   =  \alpha_R  \sqrt{P_{U}} s_{U,k}  +  \alpha_R \boldsymbol{w}_{U,k}^H \boldsymbol{z}_R +   \boldsymbol{h}_{R,k}^T \boldsymbol{H}_{RB} ^{\dag}  \boldsymbol{z}_{B},
 \end{equation}
where $\boldsymbol{w}_{U,k}$ denotes the $k$-th column of the matrix $(\boldsymbol{H}_{UR}^{\dag})^H $.

 For the users, there is no ISI since  ZF precoding is used at the RS. From (\ref{eq.yR2}) into   (\ref{eq.WR1}), the received signal at the $k$-th user after removing the self-interference can be expressed as
\begin{equation}\label{eq.AyMk}
  \tilde{\boldsymbol{y}}_{U,k} = \alpha_R   \alpha_B s_{B,k}  + \alpha_R \boldsymbol{w}_{U,k}^H \boldsymbol{z}_R      + z_{U,k}.
\end{equation}

Based on (\ref{eq.yBk}) and (\ref{eq.AyMk}), both the BS and the MSs are able to estimate the designated messages from the received signal.

\section{ Asymptotic Performance Analysis}

In this section, we analyze the asymptotic performance of the  SA-ZF linear processing scheme with large antenna array at the BS and  the RS, i.e., $N_B \gg K$ and $N_R \gg K$. We first consider a fixed power case that the transmit power budgets at the BS, the RS and each user  are fixed as $P_B=E_B,P_R=E_R$ and $P_U =E_U$, where $E_B$, $E_R$ and $E_U$ are fixed numbers. Then, we will investigate two power-scaling cases to show that the users' transmit power can be significantly reduced while maintaining a good performance.

\subsection{Asymptotic Spectral Efficiency for the Fixed Power Case}

\proposition For the cTWRN with a fixed number of users  and fixed power budgets   $P_B=E_B,P_R=E_R$ and $P_U =E_U$, when the number of antennas at the BS and the RS grow  large with a fixed ratio, i.e.,$N_B \to +\infty$,    $N_R \to +\infty$,  and $\theta_{BR} = \frac{N_B}{N_R}$ is fixed,  the asymptotic signal-to-noise ratios (SNRs) at the users and the BS for the  SA-ZF linear processing scheme are given by

\begin{equation}\label{eq.SINRSAZFMS}
\gamma_{U,k}  =  \frac{N_R(\theta_{BR}-1) {\ell_{U}} \ell_{B} E_B E_R}{ K \left[   \ell_{U} (E_R +KE_U)+ (\theta_{BR}-1) \ell_{B} E_B  \right] \sigma^2},
\end{equation}
and
\begin{equation}\label{eq.SINRSAZFBS}
\gamma_{B,k}   =  \frac{ N_R (\theta_{BR}-1) \ell_{B} \ell_{U} E_R E_U}{\left[  (\theta_{BR}-1) \ell_{B}( E_R +  E_B )+  \ell_{U} K E_U \right] \sigma^2},
\end{equation}
respectively, where $k=1,\ldots,K$.

\begin{proof} First, consider the power normalizing factors $\alpha_{B}$ and $\alpha_R$.
With large antenna arrays at the BS and the RS, from (\ref{eq.FB1}), $\alpha_B$ can be determined by
\begin{equation}\label{eq.aB1}
\begin{split}
 \alpha_B &= \sqrt{\frac{E_B}{ \mathbb{E} \left\{  \text{tr} \left[ \boldsymbol{H}_{UR} \boldsymbol{H}_{UR} ^H \left( \boldsymbol{H}_{BR} \boldsymbol{H}_{BR}^{ H}  \right)^{-1}   \right] \right\} }}  \\
&= \sqrt{\frac{E_B}{K \ell_{U} \mathbb{E} \left\{ \text{tr} \left[   \left( \boldsymbol{H}_{BR} \boldsymbol{H}_{BR}^{ H}  \right)^{-1}   \right] \right\} }}  \\
&= \sqrt{\frac{(N_B-N_R) {\ell}_{B} E_B}{N_R K \ell_{U}}},
\end{split}
\end{equation}
 where in the last step we   used the identity  $\mathbb{E} \left[ \text{tr} (\boldsymbol{X}^{-1}) \right] = \frac{M}{N-M}$, when $ \boldsymbol{X}$ is a $M \times M$ central Wishart matrix with $N$ degrees of freedom \cite{Tulino}.

 From  (\ref{eq.yRSAZF}), (\ref{eq.xR}) and (\ref{eq.WR1}), we have
\begin{equation}\label{eq.aR1}
\begin{split}
 \alpha_R  &=  \sqrt{\frac{E_R}{(\alpha_{B}^2 + E_U) \mathbb{E} \left\{ \text{tr}  \left[ \left(\boldsymbol{H}_{RU} \boldsymbol{H}_{RU}^H \right)^{-1} \right] \right\}}}\\
&= \sqrt{\frac{(N_R-K) {\ell}_{U} E_R}{ K  ( \alpha_B^2 + E_{U})  }}.
\end{split}
\end{equation}

Now consider the received SNR at the $k$-th user. Note that the variance of the noise $  \boldsymbol{w}_{U,k}^H \boldsymbol{z}_R$ in (\ref{eq.AyMk}) is given by \cite{Ngo14}
\begin{equation}\label{eq.AVarMk}
\begin{split}
\mathbb{V}\text{ar}\left[\boldsymbol{w}_{U,k}^H \boldsymbol{z}_R \right] &= \sigma^2 \mathbb{E} \left[ \left( (\boldsymbol{H}_{UR}^H \boldsymbol{H}_{UR})^{-1}  \right)_{kk} \right] \\
 &=   \frac{\sigma^2}{{\ell}_{U}} \mathbb{E} \left[  \left( ( \tilde{\boldsymbol{H}}_{UR}^H \tilde{\boldsymbol{H}}_{UR})^{-1}  \right)_{kk} \right] \\
&=   \frac{\sigma^2}{K {\ell}_{U}} \mathbb{E} \left[  \text{tr} \left( ( \tilde{\boldsymbol{H}}_{UR}^H \tilde{\boldsymbol{H}}_{UR})^{-1}  \right)  \right] \\
&=   \frac{\sigma^2}{ (N_R-K) {\ell}_{U}}.
\end{split}
\end{equation}

From (\ref{eq.AyMk}), the received SNR at the $k$-user is
\begin{equation}\label{eq.A.SINRMk}
\begin{split}
\gamma_{U,k} &=  \frac{\alpha_R^2 \alpha_B^2}{\alpha_R^ 2\mathbb{V}\text{ar}\left[ \boldsymbol{w}_{U,k}^H \boldsymbol{z}_R \right] + \sigma^2} \\
&= \frac{(N_R/K-1)(\theta_{BR}-1) {{\ell}_{U}} \ell_{B} E_B E_R}{ \left[ \ell_{U} (  E_R   +      K  E_{U}) +  (\theta_{BR}-1) \ell_{B} E_B  \right] \sigma^2}.
\end{split}
\end{equation}
Using the fact that $N_R/K - 1 \simeq N_R/K$ for large $N_R$, we obtain the result in (\ref{eq.SINRSAZFMS}).

From (\ref{eq.yBk}), the received SNR of the $k$-stream at the BS is
\begin{equation}\label{eq.ASINRBk}
\gamma_{B,k}=  \frac{\alpha_R^2 P_U}{\alpha_R^ 2 \mathbb{V}\text{ar}\left[ \boldsymbol{w}_{U,k}^H \boldsymbol{z}_R \right] + \mathbb{V}\text{ar}\left[ \boldsymbol{h}_{R,k}^T \boldsymbol{H}_{RB} ^{\dag}  \boldsymbol{z}_{B} \right]}.
\end{equation}
With large antenna arrays, the variance of the noise term $\boldsymbol{h}_{R,k}^T \boldsymbol{H}_{RB} ^{\dag}  \boldsymbol{z}_{B}$ can be calculated as
\begin{equation}\label{eq.AVar2}
\begin{split}
\tilde{\sigma}_{B,k}^2 &=  \sigma^2 \mathbb{E} \left\{  \text{tr} \left[  \left(\boldsymbol{h}_{R,k}^T \boldsymbol{H}_{RB} ^{\dag}  \right)^H   \boldsymbol{h}_{R,k}^T \boldsymbol{H}_{RB} ^{\dag} \right] \right \} \\
&= \sigma^2 \mathbb{E} \left\{ \text{tr} \left[ \left( \boldsymbol{H}_{RB} ^H \boldsymbol{H}_{RB}  \right)^{-1}  \boldsymbol{h}_{R,k}^*  \boldsymbol{h}_{R,k}^T  \right] \right \} \\
 &= \frac{N_R {\ell}_{U} \sigma^2}{(N_B -N_R) \ell_{B}}.
\end{split}
\end{equation}
Substituting (\ref{eq.AVarMk}) and (\ref{eq.AVar2}) into (\ref{eq.ASINRBk}), we have
\begin{equation}\label{eq.A.SINRBk}
\gamma_{B,k}=  \frac{(N_R-K)(\theta_{BR}-1) \ell_{B} \ell_{U} E_R E_{U}}{\left[ (\theta_{BR}-1) \ell_{B} (  E_R +  E_B) +   \ell_{U} K E_{U} \right] \sigma^2},
\end{equation}
which   can be expressed as in (\ref{eq.SINRSAZFBS}) when $N_R \gg K$.
\end{proof}

From proposition 1, we can see that the small-scale fading is averaged out when both the BS and the RS are equipped with large antenna arrays. Also, inter-stream interference diminishes with zero-forcing processing. For the cTWRN with fixed transmit power budgets, and with fixed antenna ratio $\theta_{BR}$, the received SNRs at the BS and the users increase  linearly with the number of antennas at the relay, which implies that a diversity gain of $N_R$ is achieved.

Based on the SNR expressions in  (\ref{eq.SINRSAZFMS}) and (\ref{eq.SINRSAZFBS}), the asymptotic spectral efficiency  of the cTWRN with SA-ZF linear processing is given by
\begin{equation}\label{eq.Rsum}
R_{sum} = \sum_{k=1}^K [ {\cal C}(\gamma_{B,k}) + {\cal C}(\gamma_{U,k})],
\end{equation}
where ${\cal C}(x) = \frac{1}{2} \log_2(1+x)$.

\proposition The spectral efficiency  gap between the SA-ZF scheme and the cut-set bound in the high SNR region is
\begin{equation}\label{eq.RsumGapSAZF}
\begin{split}
\Delta R_{sum}    &= \frac{K}{2}   \log_2 \left[ 1 + \frac{ \lambda (E_R + K   E_U)}{  (\theta_{BR}-1)  E_B  } \right]  \\
 & + \frac{K}{2}   \log_2 \left[ 1 + \frac{E_B}{E_R} + \frac{   \lambda  K   E_U } { (\theta_{BR}-1)  E_R}\right],
 \end{split}
\end{equation}
where $\lambda = \ell_{U}/\ell_{B}$.

\begin{proof}
When the BS and the RS are equipped with large antenna arrays, the network thought is bottlenecked by the  links between the users and the relay, and the cut-set bound is given by \cite{YangHJ12}
\begin{equation}
\begin{split}\label{eq.Rcutset}
R_{sum,cs}  &= \frac{1}{2} \log_2   |\boldsymbol{I}_{N_R} +   \boldsymbol{H}_{RU}^H \boldsymbol{Q}_{R} \boldsymbol{H}_{RU} / \sigma^{2}|  \\
& +  \frac{1}{2} \log_2 | \boldsymbol{I}_{N_R} +    \boldsymbol{H}_{UR} \boldsymbol{Q}_U  \boldsymbol{H}_{UR}^H  / \sigma^{2}|,
\end{split}
\end{equation}
where $\boldsymbol{Q}_{R}$  and $\boldsymbol{Q}_{U}$   are the corresponding
signaling covariance matrices. In the high SNR regime, It is known that  equal power allocation at the  RS
is optimal \cite{YangHJ12}, i.e., $ \boldsymbol{Q}_{R} = \frac{E_R}{K} \boldsymbol{I}_K$. Also, as
the users can not cooperate, $ \boldsymbol{Q}_{U}$ is given by $ \boldsymbol{Q}_{U} = E_U \boldsymbol{I}_K$. Then using the identity $\det(\boldsymbol{I} + \boldsymbol{A} \boldsymbol{A}^H) = \det(\boldsymbol{I} + \boldsymbol{A}^H \boldsymbol{A})$, and the approximation that $\boldsymbol{H}_{RU} \boldsymbol{H}_{RU}^H =\boldsymbol{H}_{UR}^H \boldsymbol{H}_{UR} \simeq \ell_{U} N_R \boldsymbol{I}_{K}$   for large $N_R$, the spectral efficiency   cut-set bound at high SNR can be written as
\begin{equation}\label{eq.A.RsumCS}
R_{sum,cs} = \frac{K}{2}   \log_2 \left(   \frac{N_R \ell_{U} E_U}{\sigma^2}  \right) + \frac{K}{2}   \log_2 \left(  \frac{N_R \ell_{U} E_R}{K\sigma^2}  \right).
\end{equation}

From (\ref{eq.SINRSAZFMS}), (\ref{eq.SINRSAZFBS}), (\ref{eq.Rsum}) and (\ref{eq.A.RsumCS})   , we can obtain the   performance gap in (\ref{eq.RsumGapSAZF}).
\end{proof}

From proposition 2, we conclude that the SA-ZF scheme  is able to closely approach the sum capacity of the cTWRN under certain conditions. Specifically, if $(\theta_{BR}-1)  E_B \gg  \lambda (E_R + K E_U)$, $E_R \gg E_B $ and $ (\theta_{BR}-1)  E_R \gg \lambda  K E_U $, then from (\ref{eq.RsumGapSAZF}), we have $ \Delta R_{sum}   \to 0$. This result suggests that the spectral efficiency of the SA-ZF linear processing scheme can approach the cut-set bound under the  condition that the transmit power of the RS is much higher than the BS and the users ($E_R \gg E_B \gg KE_U$), the number of antennas at the BS is much larger than that at the RS ($\theta_{BR} \gg 1$), and the path-loss between the BS and the RS is smaller than that between the users and the RS, such that $(\theta_{BR}-1)  E_B \gg \lambda E_R $. Note that these conditions can be satisfied in practical systems. This result also implies that the noise propagation in the AF cTWRN can be greatly suppressed when both the BS and RS are equipped with large antennas arrays. In contrast, it is known that there is at least 3 dB performance loss in the AF cTWRN when the BS and RS are equipped with finite number of antennas \cite{Fang14}.

\subsection{Power Scaling Laws}
In the following, we investigate two power-scaling cases for the cTWRN with SA-ZF relaying protocol: I) The transmit power of the BS and the RS are fixed, while the transmit power of each user is inversely proportional to the number of antennas at the relay, i.e., $P_B=E_B, P_R=E_R$, and $P_U = \frac{E_U}{N_R}$, where $E_B,E_R$, and $E_U$ are fixed; and II) The transmit power of the RS is fixed,  while the transmit power of the BS  and each user is inversely proportional to the number of antennas at the relay, i.e., $P_B=\frac{E_B}{N_R}, P_R= {E_R} $, and $P_U = \frac{E_U}{N_R}$.

For power-scaling case I, following a similar approach as in Proposition 1, the asymptotic SNRs at the users and the  BS are  given by
\begin{equation}\label{eq.SINRSAZFMk1} 
\gamma_{U,k}^{\text{\text{I}}}  =  \frac{ N_R (\theta_{BR}-1) {\ell_{U}} \ell_{B} E_B E_R}{ K \left[   \ell_{U} E_R + (\theta_{BR}-1) \ell_{B} E_B  \right] \sigma^2}, 
\end{equation}
and
\begin{equation} \label{eq.SINRSAZFBk1}
\gamma_{B,k}^{\text{\text{I}}}  =  \frac{  \ell_{U} E_R E_U}{\left(    E_R +   E_B  \right) \sigma^2},  
\end{equation}
$k=1,\ldots,K,$, respectively. In addition, the corresponding spectral efficiency  gap between the SA-ZF scheme and the cut-set bound in the high SNR region is
\begin{equation}\label{eq.RsumGapSAZF1}
\begin{split}
\Delta R_{sum}^{\text{I}}    & = \frac{K}{2}   \log_2 \left( 1 + \frac{E_B}{E_R}  \right)   \\
& + \frac{K}{2}   \log_2 \left[ 1 +  \frac{   \lambda    E_R } { (\theta_{BR}-1)  E_B}\right].
\end{split}
\end{equation}

From (\ref{eq.SINRSAZFMk1}) and (\ref{eq.SINRSAZFBk1}), we can see that for power-scaling case I, the received SNR at the $k$-th user  is proportional to the number of antennas at the RS when $\theta_{BR}$ is fixed, while the SNRs at the BS are constants independent of $N_B,N_R$, and the number of users $K$. Hence, for a fixed number of users, the achievable rate in the downlink increases as $N_R$ becomes larger, while the uplink transmission  rate is a constant for each user. This result is attractive for practical cellular networks, since a higher data rate is demanded in the downlink while a low transmit power in the uplink can  extend the lifetime of mobile stations. From (\ref{eq.RsumGapSAZF1}), we conclude that the SA-ZF under power-scaling case I can also approach the corresponding cut-set bound, provided that $E_R \gg E_B$  and $(\theta_{BR}-1)  E_B \gg  \lambda    E_R$.


 For power-scaling case II,   the asymptotic SNRs at the users and the BS  are given by
\begin{equation}\label{eq.SINRSAZFMk2} 
\gamma_{U,k}^{\text{\text{II}}}  =  \frac{ (\theta_{BR}-1)  \ell_{B} E_B}{ K \sigma^2}, 
\end{equation}
and
\begin{equation}\label{eq.SINRSAZFBk2} 
\gamma_{B,k}^{\text{\text{II}}}  =  \frac{   \ell_{U}  E_U}{ \sigma^2}, 
\end{equation}
$k=1,\ldots,K, $, respectively. In addition, we can show that the corresponding spectral efficiency  gap between the SA-ZF scheme and the cut-set bound in the high SNR region is
\begin{equation}\label{eq.RsumGapSAZF2}
\Delta R_{sum}^{\text{II}}      =  \frac{K}{2}   \log_2 \left[   \frac{  \lambda N_R     E_R } { (\theta_{BR}-1)  E_B}\right].
\end{equation}

As in the fixed power case and power-scaling case I, the small-scale fading is averaged out and  the inter-stream interference also diminishes. Note that  both $\gamma_{U,k}^{\text{\text{II}}}$ and $\gamma_{B,k}^{\text{\text{II}}}$ are independent of $N_R$ in this case. The asymptotic SNRs in (\ref{eq.SINRSAZFMk2}) and (\ref{eq.SINRSAZFBk2}) show that the transmit power of the BS and the users  can be made inversely proportional to the number of relay antennas while maintaining a given  quality-of-service. Note that reducing the transmit power of BS  can help in cutting the electricity power consumption. From (\ref{eq.RsumGapSAZF2}), we see that for power-scaling case II, there will be certain performance loss as compared with the corresponding cut-set bound, and this performance gap enlarges as the   number of relay antennas grow large.

\remark From the above discussions, we see that with large antenna arrays, the small-scaling fading is averaged out, the noise propagation effect can be greatly suppressed and inter-stream interference diminishes for both the fixed power case and the two power-scaling cases. Among the three considered power cases, the fixed power case achieves the highest spectral efficiency, while the other  two power-scaling cases are attractive for practical implementation since the transmit power of the users can be made inversely propositional to the number of relay antennas while maintaining a good performance.

%
%
\section{Numerical results}

\begin{figure}[t]
\centering
\includegraphics[width=3.6  in]{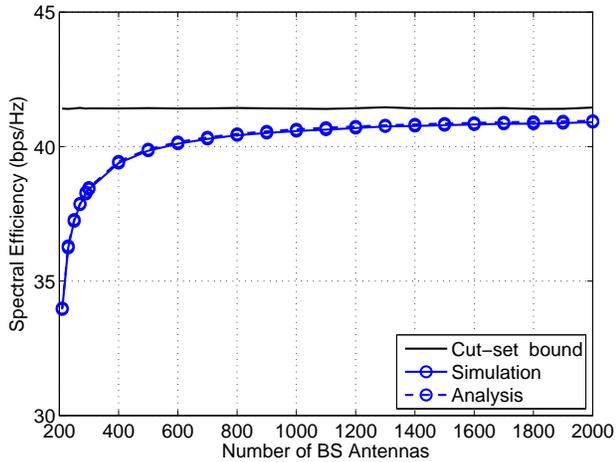}\\
 \caption{Spectral efficiency of  the cTWRN with fixed power budget, $K=3$ users,  and $ N_R=200$.}
\end{figure}

In this section, numerical results are presented to verify the analytical results.  The path-loss is chosen as $\ell_{B}=1,\ell_{U} =2^{-3}$, and each element in the small-scale fading channel matrices $\tilde{\boldsymbol{H}}_{BR}$ and $\tilde{\boldsymbol{H}}_{UR}$ is independently  complex Gaussian distributed with zero mean and unit variance. Unless otherwise specified, we assume $E_B =10$ dBm, $E_R=20$ dBm, $E_U=0$ dBm, and the noise variance $\sigma^2=-20$ dBm.

Fig. 2 shows the spectral efficiency of a cTWRN with $K=3$ users. The relay is equipped with $N_R=200$ antennas.   We also plot the cut-set bound for comparison. Note that the cut-set bound is independent of the BS-to-RS antenna ratio $\theta_{BR}$, since the network throughput is bottlenecked by the links between the relay and the users. The analytical result in (\ref{eq.Rsum}) is also plotted in the figure, and is shown to be very  accurate. It can also be seen that as the BS-to-RS antenna ratio $\theta_{BR}$ becomes large, the gap between the SA-ZF scheme and the cut-set bound decreases. When $\theta_{BR}=5$, i.e.,  the BS is equipped with $N_B = 1000$ antennas, the gap is only  0.27 bps/Hz per user, which is agreed with the result in Proposition 2.

\begin{figure}[t]
\centering
\includegraphics[width=3.5  in]{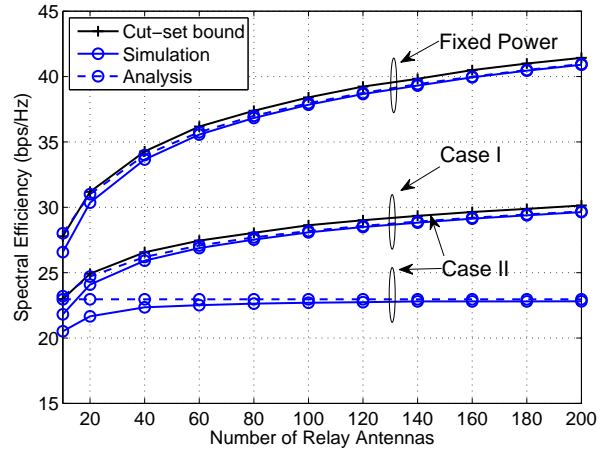}\\
 \caption{Spectral efficiency of a cTWRN with $K=3$ users, and $\theta_{BR}=10$.}
\end{figure}

The spectral  efficiency of the cTWRN versus the number of relay antennas  is shown in Fig. 3. There are $K=3$ users  and the antenna ratio is fixed as $\theta_{BR}=10$.  Both the fixed power budget case and  the two power-scaling cases are considered, and the corresponding cut-set bounds are also plotted. Note that the cut-set bounds for the power-scaling cases I and II are the same since the network throughput is bottleneckted by the links between the RS and the users. As shown in the figure, the asymptotic analysis agrees with the numerical simulation when $N_R \ge 50$, and the gap between the SA-ZF scheme and the cut-set bound is small for both the fixed power case and the power-scaling case I. For instance, the gap is within 0.2 bps/Hz per user for case I.   It can also be seen that the SA-ZF scheme with fixed power budget outperforms the other two power-scaling cases significantly, and the performance gap enlarges as the number of antennas at the relay increases. For power-scaling case II,  the spectral  efficiency  approaches to a  constant, since the received SNRs saturate  when $N_R$ is large. These results show that scaling the transmit power of the BS and each user by $1/N_R$ can still maintain a good performance.

In Fig. 4,   the spectral efficiency of the cTWRN with different number of users is plotted. The number of antennas at the relay is $N_R=200$, and the BS-to-RS antenna ratio is $\theta_{BR}=10$. For all the power  cases, the achievable spectral efficiency
increases near linearly with the number of users, and the asymptotic analysis is accurate when $K \le 40$. For $K > 40$, $K$ is comparable with $N_R$, and there is certain gap between the analytical results and the simulations.  Fig. 5 shows the corresponding energy efficiency of the SA-ZF scheme, which is defined as the spectral efficiency divided by the total consumed power. It can be seen that  when $K \le 55$,   the fixed power case achieves the highest energy efficiency. However, when $ K> 55$, the energy efficiency of the fixed power case is lower than the power-scaling case  I. This is due to the fact that the total consumed power increases rapidly as the number of users increases for the fixed power case.

\begin{figure}[t]
\centering
\includegraphics[width=3.5  in]{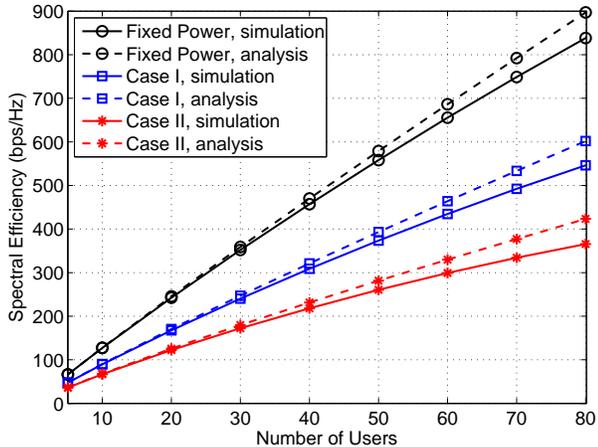}\\
 \caption{Spectral efficiency versus the number of users, $N_R=200, \theta_{BR}=10$. Sold lines denote simulations, while dashed lines denote analytical results.}
\end{figure}

%
%
\section{Conclusion}
In this paper, we investigated the performance of multiuser cellular two-way relay network with   large  antenna arrays at  the base station and the relay. We analyzed the asymptotical spectral   efficiency of the AF cTWRN with   signal-space-alignment based zero-forcing linear processing. It was shown  that by employing very large antenna arrays at the BS and the RS, the noise propagation effect  of the  AF cTWRN can be greatly reduced, the small-scale fading was averaged out, and the spectral   efficiency can be significantly improved as in point-to-point massive MIMO systems. Both theoretical analysis and numerical simulations  showed that the transmit power of the BS and the users can be made inversely proportional to the number of relay antennas while maintaining a given  quality of service.

\begin{figure}[t]
\centering
\includegraphics[width=3.5  in]{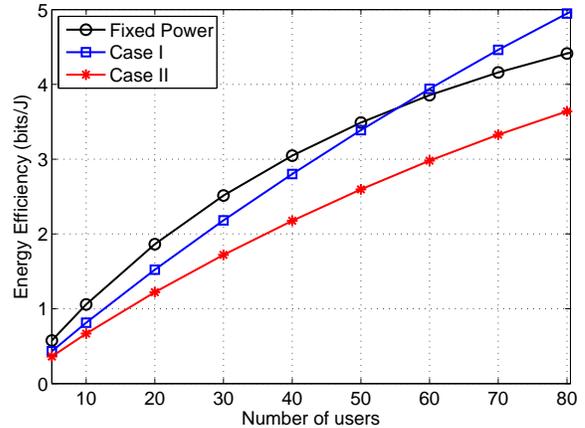}\\
 \caption{Energy efficiency versus the number of users, $N_R=200, \theta_{BR}=10$.}
\end{figure}

%
%

\end{document}